\documentstyle[11pt,paspconf,psfig]{article}

\begin{document}

\title{Dark Matter and Metal Abundances in Elliptical Galaxies
from X-ray Observations of the Hot ISM}

\author{Michael Loewenstein\affil{NASA/GSFC, Code 662,
Greenbelt, MD 20770, USA}}

\begin{abstract}
I review the results of recent analysis and interpretation of
X-ray observations of elliptical galaxies, focusing on 
elemental abundances and
dark matter. The hot ISM
is characterized by subsolar Fe abundances and solar Si-to-Fe ratios; and, I
compare these with stellar abundances and discuss 
implications of these measurements. From models constructed to explain 
X-ray temperatures and their correlation with optical properties
in a complete sample of ellipticals, I
demonstrate the ubiquity of dark matter in $L>L_*$
galaxies, present limits on dark matter structural
parameters, and discuss the scaling of dark halos
with optical luminosity.
The mass-to-light ratio within $6R_e$ has a universal value, 
$M/L_V\approx 25h_{80}$M$_{\odot}$/L$_{V_\odot}$,
that conflicts with the 
simplest extension of CDM theories of large
scale structure formation to galactic scales.
\end{abstract}


\section{Introduction and Overview}

In this contribution I review the properties of the
hot, X-ray emitting gas in elliptical galaxies. The
investigation of elliptical galaxies using X-ray observations is a
less mature and more volatile field than its radio and optical counterparts; 
however, X-ray studies provide unique and complementary insights
into the nature of these systems. I concentrate on
two topics of relevance to the subject of this conference, 
star formation in early-type galaxies.

The first topic is the nature of the X-ray emission, 
focusing on the hot gas metallicity.
The mass,
distribution, and relative abundances of metals
constrain the enrichment and, therefore, the star formation
history of these galaxies.
The complicated nature of optical abundance
studies emphasizes the value of the complementary
method of X-ray spectroscopy of hot
gas that originates as stellar mass loss.
In this portion of the review,
I explain and evaluate the ``standard''
model for the X-ray emission from elliptical galaxies, and review
the low abundances in elliptical galaxy hot interstellar media 
derived using such models.
I also compare X-ray and optical
abundances, summarize measurements of the Si-to-Fe ratio in the
hot gas,
and discuss some of the implications of the observed abundances.
Many of the results I discuss here are based on the work of
Kyoko Matsushita (Tokyo Metropolitan University),
Hironori Matsumoto (Kyoto University) and, especially,
Richard Mushotzky (NASA/GSFC).

The second part of this review is a summary
of a recently
completed project, in collaboration with Ray White 
(University of Alabama), on
the existence and
properties of dark matter halos in the population of bright
elliptical galaxies. 
Dark matter is a determining factor in the
feedback processes that occur during the star formation epoch
and are responsible for such correlations as the color-magnitude 
relation.
I review our modeling methods and
assumptions and summarize the following 
results: (a) a demonstration of the 
ubiquitousness of dark matter in ellipticals, (b) how the dark halo properties
must scale with optical properties to
match the observed X-ray/optical correlations, and (c)
implications for galaxy formation and cosmology. 

\section{X-ray Emission from Elliptical Galaxies -- General Considerations}

Surprisingly -- since it was expected that
galactic winds would drive out most gas (Mathews \& Baker 1971) -- 
the {\it Einstein} Observatory 
discovered that many elliptical galaxies are bright in X-rays, with 
luminosities up to $\sim 10^{42}$ erg s$^{-1}$
(Fabbiano 1989). The X-ray luminosity is a 
steep function of optical luminosity ($L_X\propto L_{opt}^{\sim 2.5}$), and
X-ray studies are highly biased towards the optically brightest
systems, many of which are in the Virgo cluster.

In the brightest
systems, the 
emission is dominated by $\sim 10^7$ K gas, and the gas mass within
the optical galaxy can be as high as a few percent that of the 
stars -- much less than the stellar mass loss rate
integrated over a Hubble time. 
The gross
properties of the hot gas are well explained by 
hydrodynamical models where gas is heated by stellar
motion-induced shocks and (possibly) Type Ia supernovae (SNIa), and
settles into hydrostatic equilibrium in a gravitational potential that
includes dark matter (Loewenstein \& Mathews 1987).

Some elliptical galaxies have
very extended ($>100$ kpc in radius) X-ray coronae.
This is not surprising -- a galaxy of $10^{11}$ L$_{\odot}$,
stellar $M/L=10$, and baryon
fraction 5\% has a viral radius $\sim 500$ kpc. 
It is not clear
whether the gas in these
extended halos represents a primordial baryon
reservoir or was ejected from
the galaxy during an early star forming epoch, nor is it understood why
other galaxies have compact X-ray halos.

The study of the X-ray emission from elliptical galaxies has greatly
intensified this decade due to the superior spectral and
imaging capabilities of the
{\it ROSAT} and {\it ASCA} observatories. This has led to tremendous
improvements in the accuracy and extent of derived gas density and
temperature profiles, as well as the first significant sample of
accurate gas abundances. 
These new observations are the foundation of the insights into 
elliptical galaxy structure and evolution that I now discuss.

\section{The ``Standard'' Model}

{\it ASCA} spectra can generally be decomposed into soft and hard
components (Matsumoto {\it et al.} 1997, Matsushita 1997).
The soft component originates in the hot (0.3--1.2 $10^7$ K) ISM, and
shows a wide range of X-ray-to-optical flux ratios and X-ray extents for
any optical luminosity. 
The hard component is roughly co-spatial with the optical galaxy, and
scales linearly 
with optical luminosity with a relative
normalization and spectrum
consistent with measurements of
the integrated emission from low mass X-ray binaries in spiral galaxy
bulges (although some galaxies appear to have enhanced
hard emission from a spatially unresolved nucleus). 
I will refer to this two-component model as the ``standard'' model, since
it is the simplest model
that describes the data.
Since
abundance uncertainties become large as the hard component begins
to dominate
and emission line equivalent widths 
are diluted, accurate abundances are
derivable only in
gas-rich ellipticals, of which there
are about 20 in 
the {\it ASCA} archive. This includes galaxies with both extended and
compact X-ray morphologies. 

\section{Hot Gas Abundances in the Standard Model}

Figure 1 shows a plot of abundance versus temperature 
derived from {\it ASCA} spectra
extracted from the inner $5R_e$ 
using the standard model.
The soft component is modeled using the Raymond-Smith thermal
plasma emission code
with abundances fixed at their solar photospheric ratios
(abundance of Fe relative to H
$4.68\ 10^{-5}$ by number). The abundances --
essentially the Fe abundance as X-ray spectra 
at these temperatures are dominated by Fe L emission lines --
range from about 0.1 to 0.7 solar. Since it is 
usually assumed that abundances of the mass-losing stars that
are the origin of the hot gas are supersolar, these may seem surprisingly
low. Moreover, Type Ia supernovae exploding at a 
rate $R_{SNIa}$ SNU (1 SNU $=1$ SN per $10^{10}$ 
L$_{B_\odot}$ per 100 yr)
should further enrich the hot gas by
an additional $\sim 25R_{SNIa}$ solar -- or $\sim 2.5$ times solar
using the rate from Cappellaro et al. (1997) for 
$H_0=65$ km s$^{-1}$ Mpc$^{-1}$. Thus,
X-ray abundances of elliptical galaxies
are $>5$ times less than what might
naively be expected. 

\begin{figure}[htbp]
\centerline{
\psfig{file=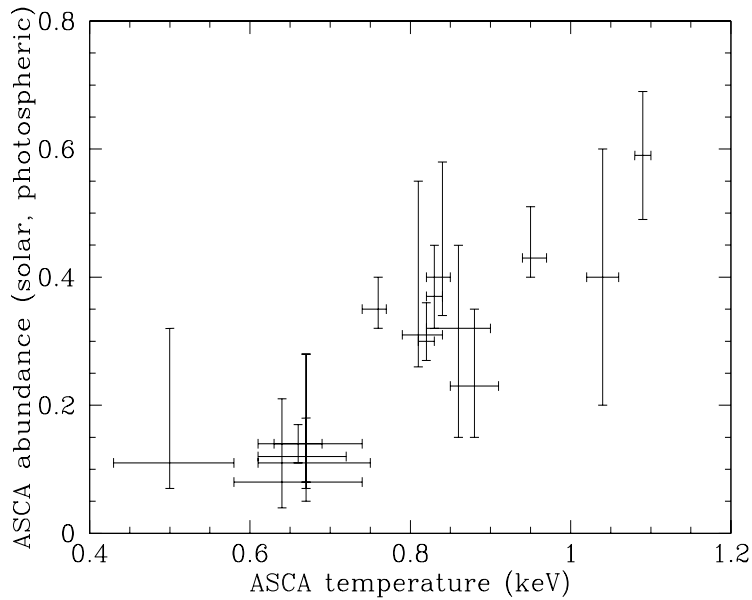,width=4.0in,height=3.2in,angle=-90,clip=}}
\caption{Hot gas metal abundance, assuming solar photospheric
ratios, versus temperature --
mostly adapted from Matsushita (1997).}
\end{figure}

\section{X-ray/Optical Metallicity Comparison Revisited}

\subsection{X-ray Advantages and Disadvantages}

Physical quantities such as hot gas abundances and temperatures are
derived from model fitting of X-ray spectra.
For abundance studies,
a great advantage of X-ray spectroscopy is that --
given sufficient signal-to-noise, spectral resolution, bandpass,
and knowledge of the 
the important atomic transitions -- emission line strengths provide 
{\it direct} measurements of elemental abundances. This
is not the case for optical abundances
(``the intensity of $Mg_2$ does not simply correlate with the abundance 
of Mg''; Tantalo, Bressan, \& Chiosi 1998).

{\it ASCA} has those qualities required for
abundance determinations of the X-ray brightest elliptical galaxies.
However, there are limitations. The bandpass
and energy resolution are insufficient 
to obtain
clean measurements of many elements; 
the atomic parameters for some
of the prominent Fe emission features are uncertain; spectra can be
complicated by multiple components not spatially separable due
to the limited {\it ASCA} angular resolution. 
``Contamination''
by SNIa explosions or accretion of intergalactic gas 
may complicate the comparison with stellar abundances. However the 
low measured hot gas Fe measurements argues against the former, while the  
lack of an anti-correlation between hot gas metallicity and X-ray luminosity 
apparently rules out the latter.

One can obtain fairly accurate Fe abundances for 20 or
so galaxies, Si abundances for about half that number, and 
occasionally some constraints on O, Mg, and S.
A comparison with the optical results
seems to be meaningful if we focus on the X-ray emission from the
optical region of the galaxy.

\subsection{Is There Evidence of an X-ray/Optical Discrepancy?}

Measurements of the nuclear $Mg_2$ index imply that
elliptical galaxies have supersolar abundances only if one assumes
(a) that metallicities are constant with
radius, (b) that abundances have
solar ratios, and (c) that
all stars were formed 
a Hubble time ago. 
For a meaningful 
comparison with the X-ray abundances, one
needs to estimate a globally averaged stellar Fe abundance. Negative abundance 
gradients indicate that the average
metallicity is typically a factor of two below the central value
(Arimoto et al. 1997), and Fe is
underabundant, typically by an additional factor of two, 
relative to Mg (Worthey, Faber, \& Gonzalez 1992). 
These factors bring the optical and
X-ray Fe abundances into fair agreement. This is illustrated by 
the {\it ASCA} Fe abundance profile for NGC 4636 
(Mushotzky et al. 1994, Matsushita et al. 1998) shown
in Figure 2, where a comparison is made
with the extrapolated optical $Mg_2$ profile
(Davies, Sadler, \& Peletier 1993) converted to [Fe/H]
assuming three separate values of [Mg/Fe]
(Matteucci, Ponzone, \& Gibson 1998). There is no clear
discrepancy once the effects of gradients and non-solar abundance ratios
are properly accounted for.

\begin{figure}[htb]
\begin{minipage}[t]{65mm}
\centerline{ 
\psfig{file=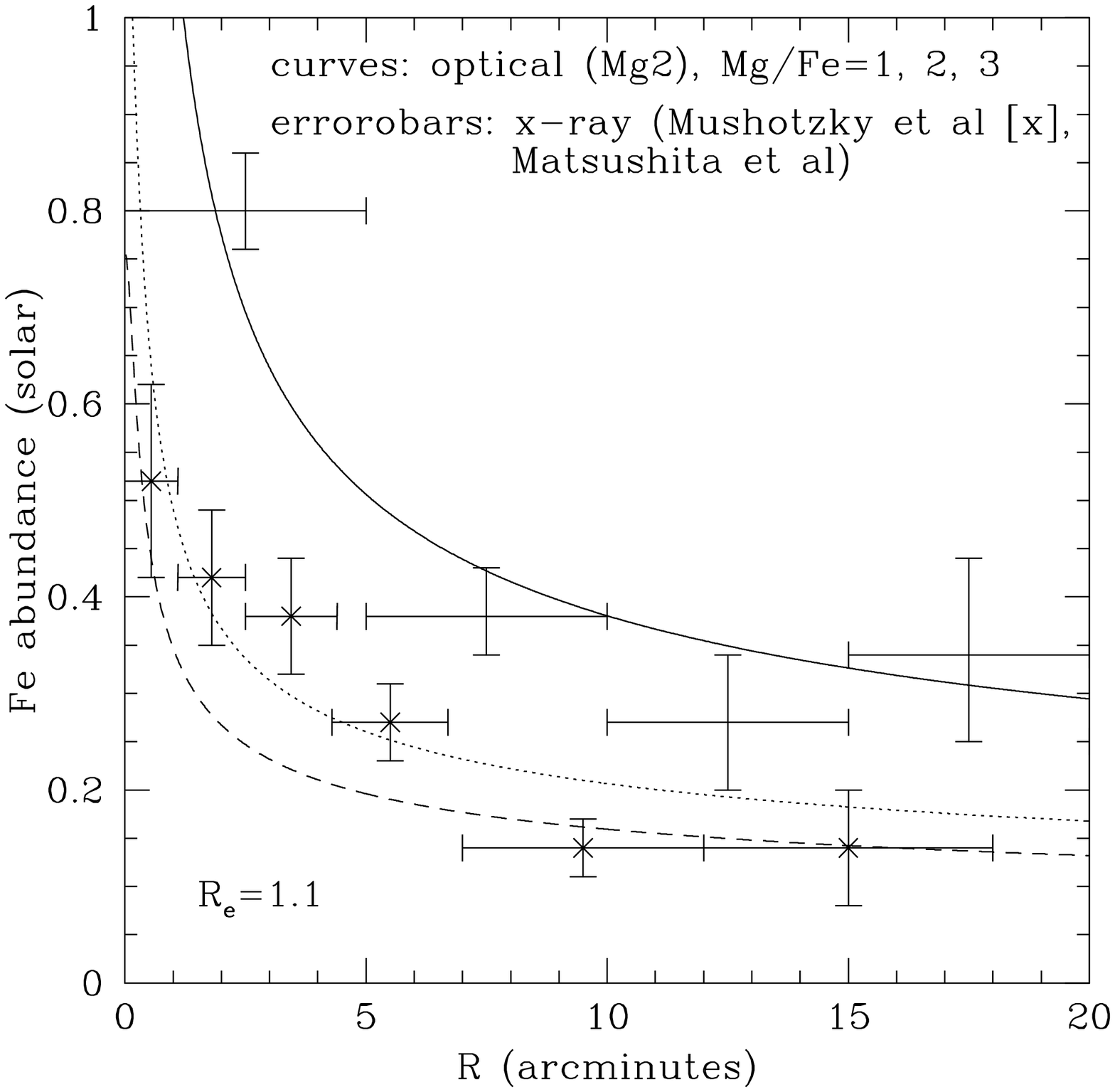,width=2.5in,height=2.5in,clip=}}
\caption{Hot ISM Fe abundance 
compared with the extrapolated optical estimates with
Mg/Fe $=1$ (solid curve), 2 (dotted curve),
and 3 (dashed curve) times solar.}
\end{minipage}
\hspace{\fill}
\begin{minipage}[t]{65mm}
\centerline{
\psfig{file=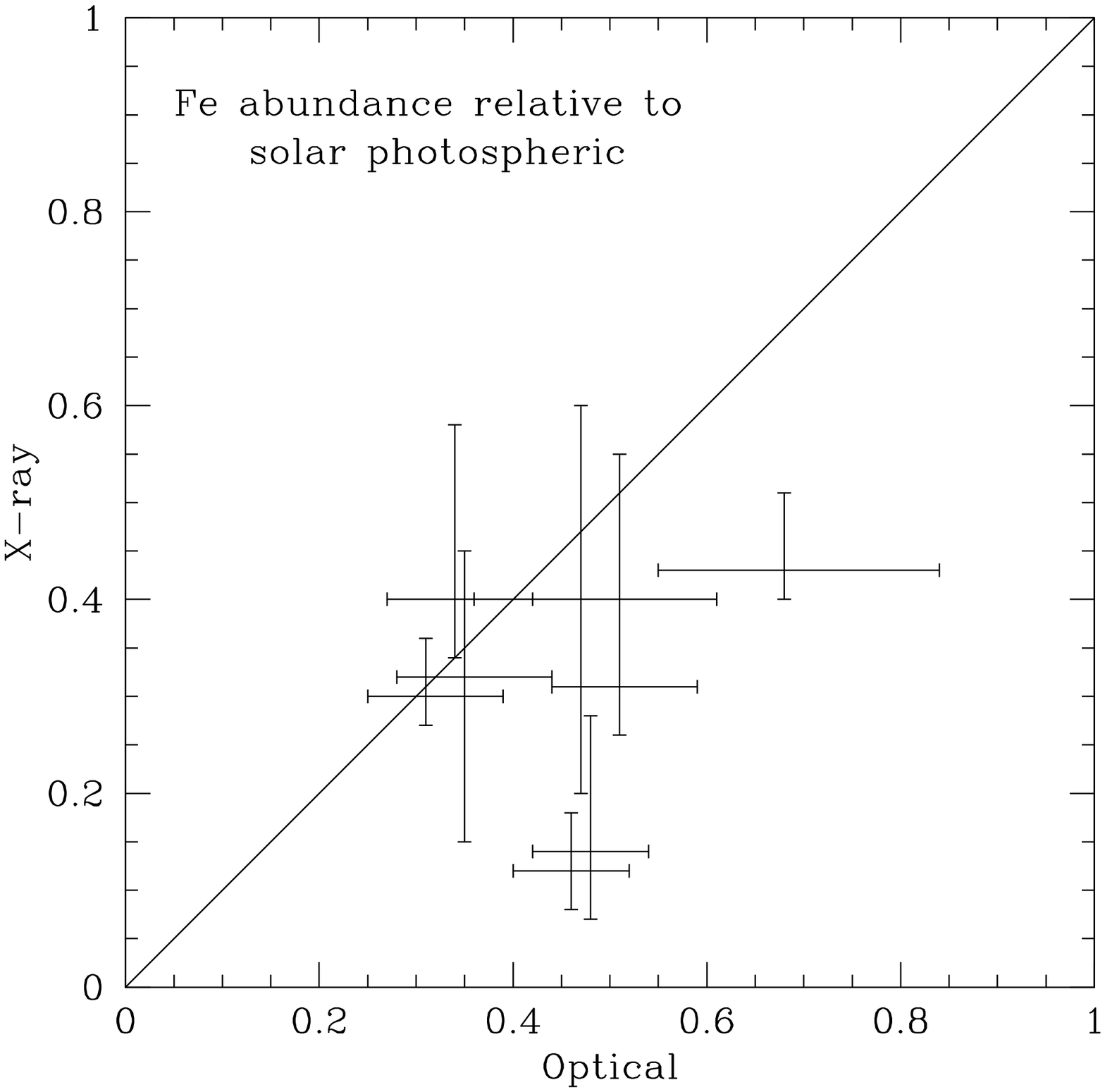,width=2.5in,height=2.5in,clip=}}
\caption{X-ray versus optical global Fe abundance.}
\end{minipage}
\end{figure}

Additional complications have emerged from
recent work in this field.
Balmer emission line measurements indicate that
many elliptical galaxies have undergone star formation relatively
recently, compromising the simple conversion from
a single line index to metallicity.
Scott Trager has kindly provided me with {\it very preliminary}
estimates, based on work underway in collaboration with
J. Gonzalez, S. Faber, and D. Burstein
that accounts for the effects of differences in
stellar population and
non-solar abundance ratios,
of the Fe abundance at the half-light radius ($R_e$)
that I have converted to a global average. There is an overlap of
eight galaxies with the gas-rich {\it ASCA} sample; and,
there is generally a rough consistency (Figure 3).
The (unweighted) average optical
Fe abundance is $\sim 0.45$ solar compared to 
$\sim 0.3$ solar for the hot gas, with
the offset dominated by two extreme gas-underabundant systems.
The ``optical/X-ray abundance discrepancy'' is
primarily an artifact of abundance gradients
and non-solar abundance ratios, greatly diminished
once the proper comparison --
of average measurements of
the same element over the same aperture -- is made. 

\subsection{Is the Standard Model Correct?}

The fair
consistency of optical and X-ray abundance determinations mitigates one
of the primary objections to the
adequacy of the simple two-component ``standard''
model. 
Inaccuracies in our knowledge of Fe L atomic physics
(Arimoto et al. 1997), has been shown
to be no more than a 20--30\% effect
(Hwang et al. 1997, Buote and Fabian 1998a). 

A more formidable alternative to the standard model
has been constructed by Buote and Fabian (1998b).
They found
that the best fit to {\it ASCA} data often consists of a two-temperature
plasma, with the
secondary component having a temperature of 1.5--2 keV,
and that the abundances in such fits are systematically higher by
about a factor of two compared
to the standard model. However, we have found
that He-like to H-like Si
line ratios are in precise agreement with the predictions of
the standard model (Figure 4; Mushotzky \& Loewenstein 1998),
although {\it ASCA} spectra are not of
sufficient quality to generally and unambiguously
rule out the Buote and Fabian two-phase model.

A final argument for the correctness of the standard model comes from
the demonstration that, in NGC 4649 and NGC 4472, the mass profile
obtained from the hot gas temperature in single phase fits is in perfect
accord with the mass determined optically (Brighenti \& Mathews 1997).

\begin{figure}[htbp]
\centerline{
\psfig{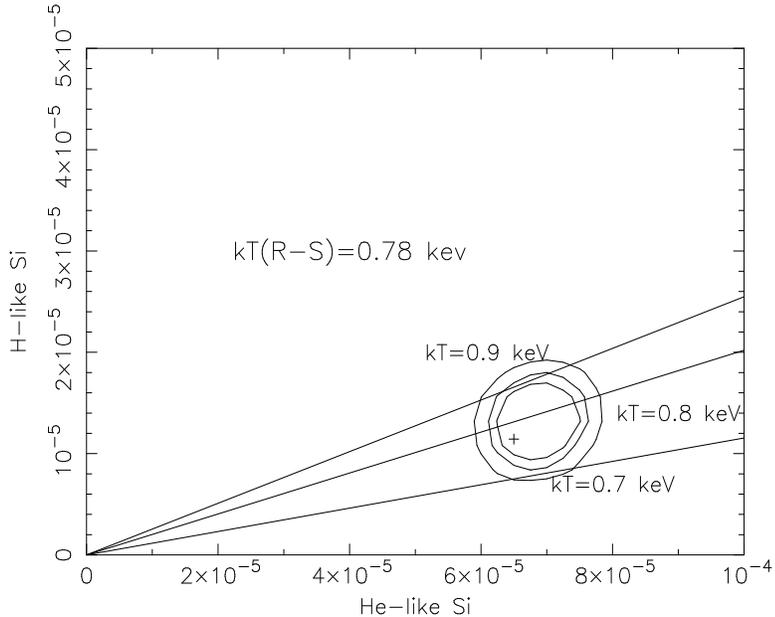}}
\caption{68, 90, and 99\% confidence contours
for H- and He-like Si line strengths (in photons cm$^{-2}$ s$^{-1}$) in
the elliptical galaxy NGC 4636.
The solid lines show the ratios
expected in three single-temperature thermal plasma models.
The temperature derived from
global spectral fitting
is in precise agreement with
the line diagnostic value.}
\end{figure}

\section{Si-to-Fe Ratio in the Hot Gas}

Elemental abundance ratios provide constraints on the primordial
IMF and relative numbers of Type Ia and Type II supernovae.
Renormalized to the meteoritic Fe abundance, the 
Si-to-Fe ratio lies between 0.5 and 1.5 times solar (Figure 5).
This is lower than the
Mg-to-Fe ratio derived from nuclear optical spectra, and is more in line with
values measured from the X-ray spectra of intragroup media.
This implies that 
either the $\alpha$-to-Fe enhancement
is a phenomenon restricted to the inner ($<R_e$) galaxy, or that
the Si-to-Mg ratio is subsolar. Evidently, intracluster media 
(Loewenstein \& Mushotzky 1996) and
elliptical galaxy cores have the enhanced $\alpha$-to-Fe 
elemental ratios characteristic
of rapid high mass star formation where enrichment is dominated by Type II
supernovae, while groups and the
outer regions of elliptical galaxies tend toward the solar supernovae
mix where a larger fraction of Fe originates in SNIa.

\begin{figure}[htbp]
\centerline{
\psfig{file=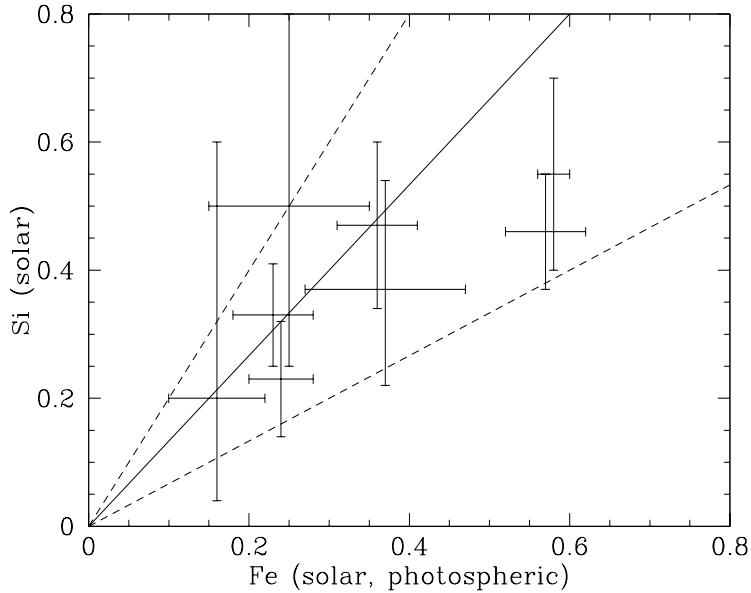,width=4.0in,height=3.2in,angle=-90,clip=}}
\caption{Si versus Fe abundance in the hot X-ray emitting gas.
The solid line denotes Si:Fe in the ratio 1:1, while
the broken lines denote the ratios 3:2 and 1:2
with respect to the (meteoritic) solar ratio.}
\end{figure}

The Si abundance provides a robust
upper limit on the effective SNIa
rate that is consistent with that derived using Fe. The 
conservative (assuming that all of the Si in the hot gas originates
from SNIa)
limit is typically $\sim 0.03$ SNU --
about four times lower than the
recent estimate of Cappellaro {\it et al.} (1997).

\section{Implications of Low Abundances}

A globally averaged Fe abundance in elliptical galaxies of
about half-solar is in accord with
optical and X-ray spectroscopic measurements, as well as with 
predictions of
chemically consistent evolutionary
models (M\"oller, Fritze-v. Alvensleben, \& Fricke 1997).
As this is only slightly
higher than the ICM Fe abundance, and the ICM dominates the
cluster baryon mass,
there is considerably more Fe
in the ICM than is locked
up in cluster galaxy stars. This implies one of the following.

(1) If the stellar and ICM metals come from the same SNII-enriched
proto-elliptical galaxy gas, then 50--90\% of the original galaxy mass was
lost and much of the ICM is not primordial but was ejected
from galaxies.
(2) However, the actual mass of material directly associated
with the SNII ejecta is much less significant.
Selective mass-loss of nearly pure SNII ejecta would enable expulsion 
of much of the metals while retaining most of the baryonic mass.
There is both 
observational (Kobulnicky \& Skillman 1997) and theoretical 
(Mac Low \& Ferrara 1998) evidence
for super-enriched outflows in dwarf galaxies that may serve as
analogues of pre-merger elliptical galaxy sub-units.

(3) It is possible that the ICM enrichment originates in
some other source. The most plausible candidates are dwarf galaxies, but
Gibson \& Matteucci (1997) have shown that this scenario is not consistent
with the color-magnitude relation in these systems. Therefore,
one may have to appeal to a population of 
dwarf galaxies that destroy
themselves in the process of enriching the ICM. 

Models of the chemical evolution of elliptical
galaxies are often tuned to produce supersolar 
stellar abundances, reproduce the
color-magnitude diagram, explain ICM enrichment, etc.
They also tend to predict high ISM metallicities
(e.g., Matteucci \&  Gibson 1995).
Re-evaluation of these models in light of the
downward revision of elliptical galaxy metallicities may be in order.

\section{Dark Matter in Elliptical Galaxies: Background and Motivation}

I now turn to the second main topic of this review, dark matter in
elliptical galaxies.
There is a strong consensus that dark matter dominates the mass content
of spiral galaxies,
galaxy groups and
clusters, and the universe as a whole. Although
traditionally less forthcoming and more controversial, 
evidence for dark matter in elliptical galaxies has 
rapidly accumulated in recent years from improved
stellar dynamical data and modeling,
gravitational lensing
observations, and high-quality X-ray images and spectra from
the
{\it ROSAT} and {\it ASCA} satellites. 
For example, the extended flat hot gas temperature profiles
measured using {\it ASCA} (Matsushita 1997) are analogous
to flat HI rotation curves in spiral galaxies as indicators
of the presence of massive dark matter halos.

Although the case for dark matter in some ellipticals
is now overwhelmingly strong, we (Loewenstein \& White 1998) were
motivated by published measurements of X-ray temperatures in a
complete optically selected 
sample (Davis \& White 1996), to  attempt to
answer the following more general
questions:
(1) Do bright elliptical galaxies have dark matter halos {\it in general}?
(2) How do the dark halo properties scale with optical luminosity?

\section{Modeling and Assumptions}

The primary diagnostic observable in this work is
the ratio of stellar to hot gas temperatures,
$\beta_{\rm spec}\equiv
\mu m_p\sigma^2/k\langle T\rangle$, where  
$\mu m_p$ is the mean mass per particle,
$\sigma$ the
projected central optical velocity dispersion, and
$\langle T\rangle$ the globally ({\it i.e.}, over
$6R_e$) averaged
gas temperature. From the fundamental plane relations and
virial theorem for the the gas, it follows that
$\beta_{\rm spec}$ is an excellent diagnostic of
the total mass-to-light ratio. 
The following 
characterize the ``$T$--$\sigma$'' relation, and must
be reproduced by any successful model of the dark matter in
elliptical galaxies: 
(1) $\beta_{\rm spec}<1$ 
(the gas is always hotter than the stars, typically by factors of 1.5--2),
and (2) $\langle T\rangle\propto \sigma^{1.45}$ or
$\beta_{\rm spec}\propto \sigma^{0.55}$ (Davis \& White 1996).

(1) Stars and gas are assumed to be in hydrostatic
equilibrium in a spherically symmetric gravitational potential.
(2) Stellar orbits are assumed to vary monotonically
from isotropic at the center to radial at infinity.
(3) Stellar density profiles and scaling relations
are determined by {\it HST} observations (Faber et al. 1997)
and the
fundamental plane.
(4) The ``NFW'' dark-matter parameterization 
(Navarro, Frenk, \& White  1997) is adopted,
\begin{equation}
\rho_{\rm dm}(r)\propto\left({r\over {R_{\rm dm}}}\right)^{-1}
\left(1+{r\over {R_{\rm dm}}}\right)^{-2},
\end{equation}
where $\rho_{\rm dm}$ and $R_{\rm dm}$ are the dark matter density distribution
and scale length, respectively.
$\beta_{\rm spec}$ is primarily
determined by the dark-to-luminous 
mass ratio inside the optical radius ($R_{\rm opt}$, 
defined here as $6R_e$,
the radius enclosing $\approx 90$\% of the light), and the
dark halo concentration
(the ratio of dark matter 
to stellar scale lengths). A global observable,
$\beta_{\rm spec}$ is 
not sensitive to the functional
form of the dark matter density distribution;
the choice of
equation (1) enables us to connect our results with 
numerical structure formation simulations.

\section{Dark Matter Universality and Limits}

$\beta_{\rm spec}>1.2$
for models without dark matter --- greater than in any
observed galaxy (Davis \& White 1996).
A typical
value of $\beta_{\rm spec}=0.5$ requires a dark-matter
fraction of $\approx 75$\% 
within $R_{\rm opt}$ for $R_{\rm dm}\approx R_e$.
Although
the dark matter distribution 
is not constrained in detail,
more than half of the mass within $R_e$ is 
baryonic for models with $\beta_{\rm spec}=0.5$ if $R_{\rm dm}>R_e$.
Even for
extreme stellar models,
{\it $\beta_{\rm spec}$ always exceeds $\approx 0.75$ unless ellipticals
have dark matter}. Therefore, dark halos
must be generic to $L>L_*$ elliptical galaxies.

We place lower limits
on the dark-matter scale length, $R_{\rm dm}$ --- if the 
dark matter is too concentrated 
$\sigma$ increases relative to $\langle T\rangle$, raising $\beta_{\rm spec}$.
The
minimum value of $R_{\rm dm}$ consistent with
$\beta_{\rm spec}\approx 0.5$ is $\approx 0.3R_e$ 
$\approx 2(L_V/3L_*)^{3/4}h^{-1}_{80}$ kpc, where 
$L_*\approx 1.7\times 10^{10}{h_{80}}^{-2}$L$_{V_\odot}$ (Figure 6).
We also derive upper limits on the baryon fraction, analogous 
to maximum disk models for spiral galaxies (Figure 7). 
The minimum dark matter mass fraction is  
$\approx30$--$57$\% within $R_{\rm opt}$ for
$\beta_{\rm spec}=0.4$--0.7,
and is $<$20\% within $R_e$.

\begin{figure}[htb]
\begin{minipage}[t]{65mm}
\centerline{ 
\psfig{file=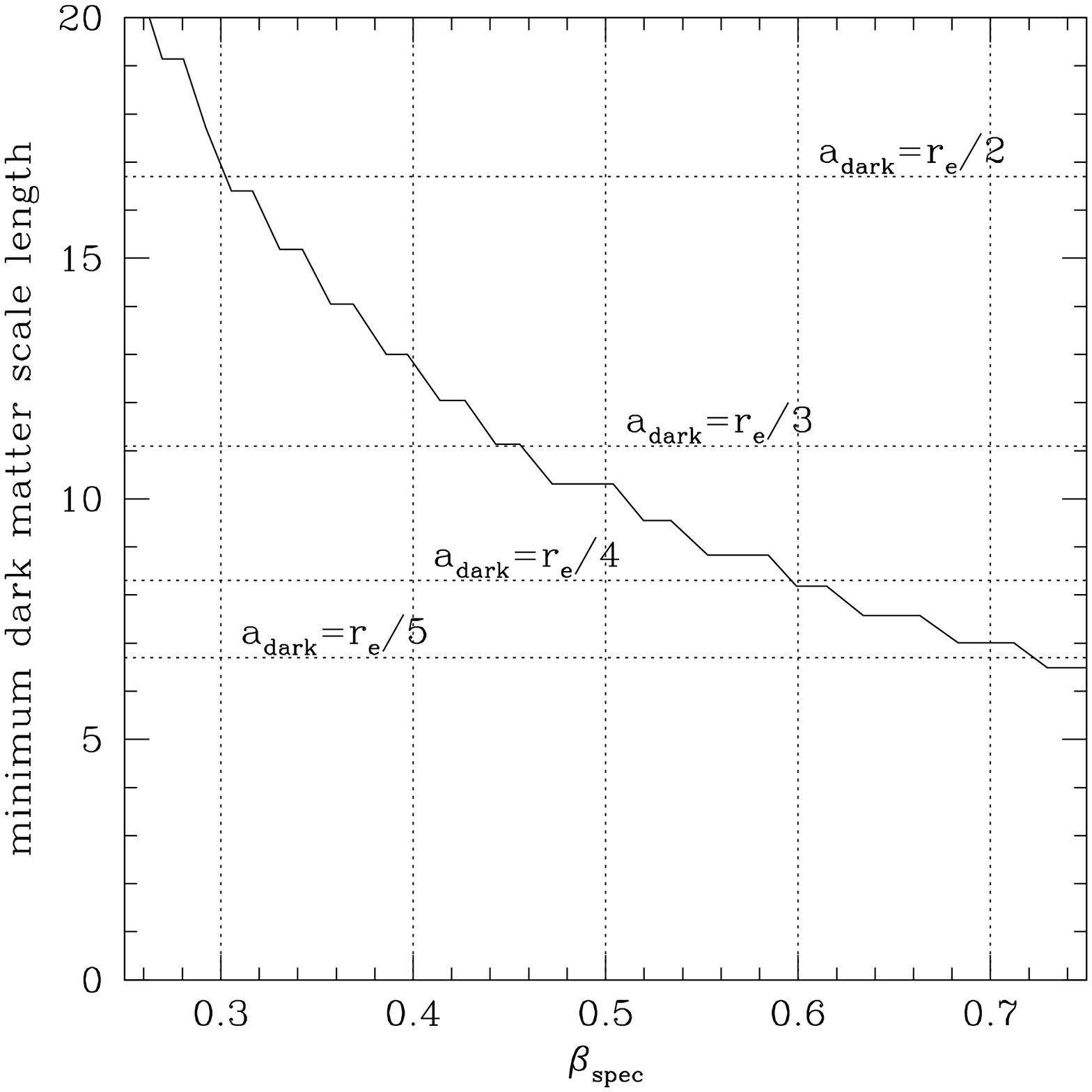,width=2.5in,height=2.5in,clip=}}
\caption{Minimum dark-matter scale length in units of the ``break radius,''
$0.03$ $R_e$.}
\end{minipage}
\hspace{\fill}
\begin{minipage}[t]{65mm}
\centerline{
\psfig{file=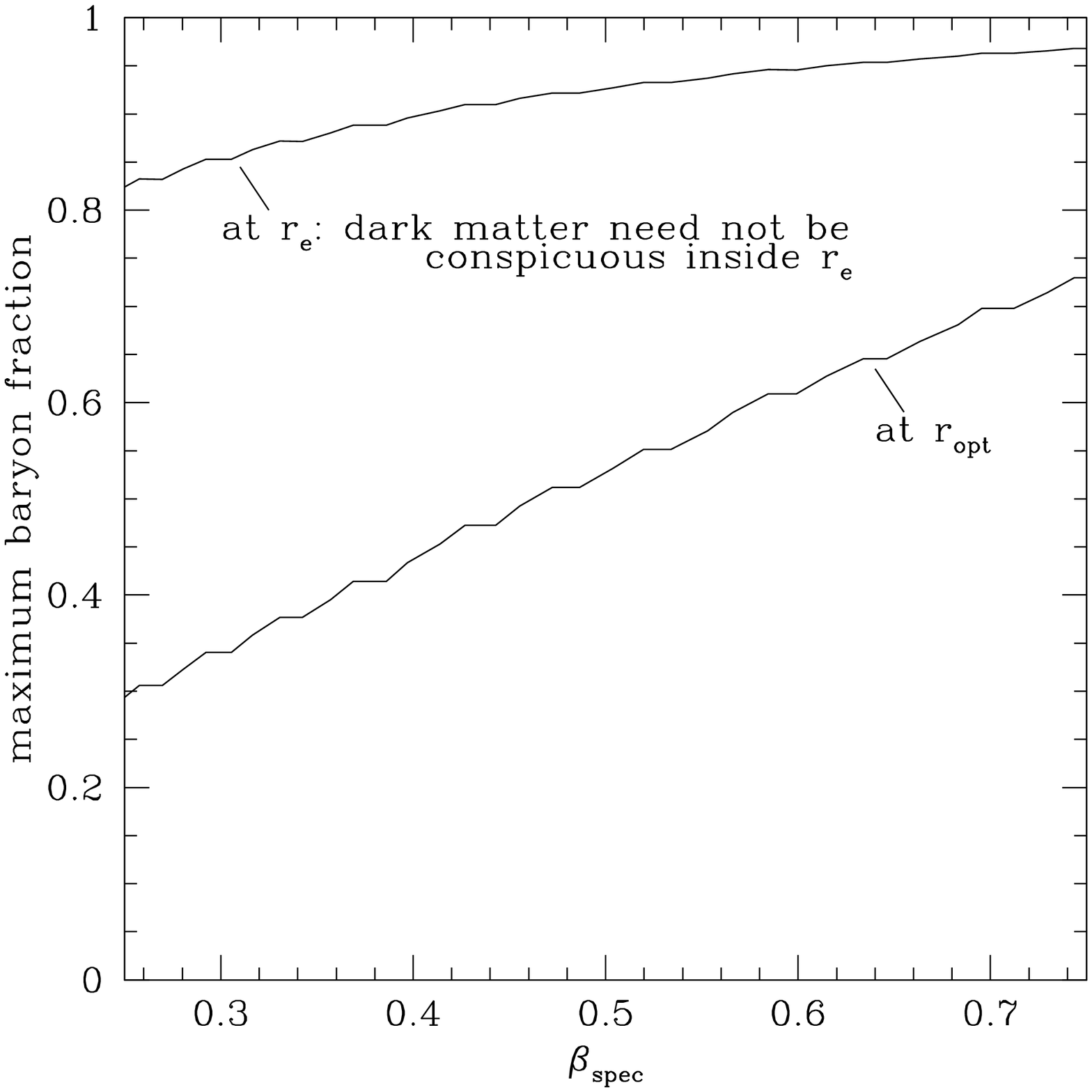,width=2.5in,height=2.5in,clip=}}
\caption{Maximum values of baryon fractions within $R_e$, and $R_{\rm opt}$.}
\end{minipage}
\end{figure}

\section{Explaining the $T$--$\sigma$ Relation}

Elliptical galaxies with the same dark-to-luminous
mass and scale length ratios have the same 
$\beta_{spec}$. As
can be inferred from the virial theorem
and fundamental plane relations,
the observed trend wherein
$\beta_{\rm spec}$ increases with increasing $\sigma$ 
(or, equivalently, with $L_V$) implies that
more luminous 
galaxies are {\it less} dark-matter dominated within $R_{\rm opt}$
(and in such a way that the total mass-to-light ratio is
nearly constant).
Extending our models to virial radius and mass scales, we
investigate what dark matter scaling
relations predict such a trend.
Two successful scenarios are the following (Figure 8).
(1) The dark matter scale length
$R_{\rm dm}$ increases weakly
with $M_{\rm virial}$ as predicted in CDM simulations, but
the baryon fraction ($f_{\rm bar}$) is
an increasing function of optical luminosity, as 
expected if smaller galaxies undergo more intense
supernova-driven mass loss during their star forming epoch
(dotted curves in Figures 8, 9, and 10).
If all ellipticals formed with the same $f_{\rm bar}$, then
the average $L>L_*$ galaxy has lost more than half its original mass.
(2) All elliptical galaxies have the same $f_{\rm bar}$
but $R_{\rm dm}$ increases {\it much more
steeply} with $M_{\rm virial}$ than in CDM models
(dashed curves in Figures 8, 9, and 10). In this case,
less luminous galaxies have relatively more dark matter within
$R_{\rm opt}$ because of a more concentrated dark-matter distribution
rather than a larger overall dark-matter fraction.
This deviation from CDM predictions of dark halo scaling
on mass scales $<10^{14}$M$_{\odot}$
could result from a relatively flat primordial fluctuation spectrum
or the effects on the
dark matter density profile from
evolution of the baryonic component.

\begin{figure}[htb]
\centerline{
\psfig{file=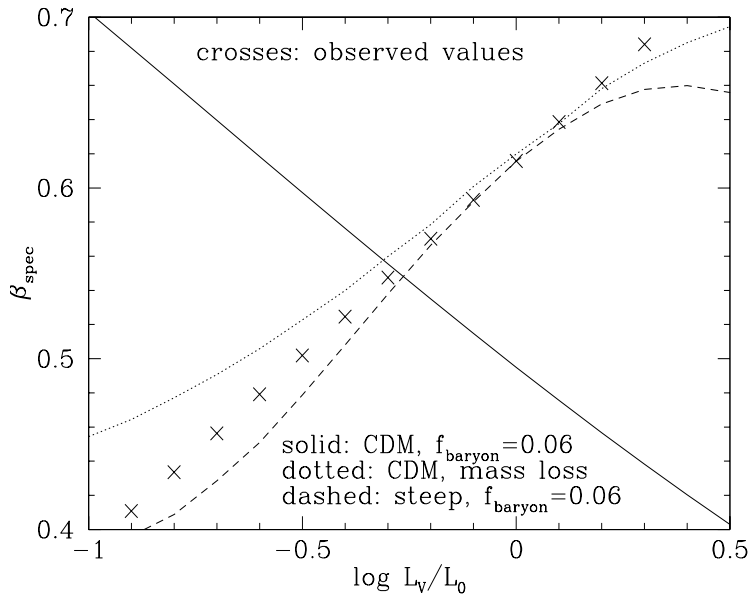,width=4.0in,height=3.2in,angle=-90,clip=}}
\caption{Observed and predicted 
correlation of $\beta_{\rm spec}$ with dimensionless
luminosity ($L_o=5.2\times 10^{10}{h_{80}}^{-2}$L$_{V_\odot}$).
Solid curve denotes constant $f_{\rm bar}$ and CDM 
dark matter concentration scaling; dotted curve has $f_{\rm bar}$ increasing
with luminosity; dashed curve has a steeper-than-CDM
scaling of concentration with dark halo mass.}
\end{figure}

{\it Models with dark halos that scale as predicted by CDM,
but with constant $f_{\rm bar}$,
badly fail to reproduce the observed $T$--$\sigma$ trend} (solid curve in
Figure 8).

\section{How Dark Matter Scales with Optical Luminosity}

For the two scenarios described above that successfully reproduce the
$T$--$\sigma$ relation, total masses within
$R_e$ and $R_{\rm opt}$ are perfectly consistent
with gravitational 
lensing results (Griffiths et al. 1996;
see Figure 9), as well as with those
from studies of
ionized gas disks as discussed by R. Morganti at this meeting.
Integrated properties within $R_{\rm opt}$
are robust (Figure 10): 
$M/L_V\approx 25h_{80}$M$_{\odot}$/L$_{V_\odot}$, 
($f_{\rm bar}\approx 0.35(L_V/3L_*)^{1/4}$).
On scales both larger and smaller than $R_{\rm opt}$, dark-matter scaling in
the two scenarios described above diverges (Figures 9 and 10).
In the constant $f_{bar}$, non-CDM scaling
scenario, dark
matter becomes increasingly important inside $R_e$ as $L_V$
decreases, becoming dominant for $L<0.6L_*$.

\begin{figure}[htb]
\begin{minipage}[t]{65mm}
\centerline{
\psfig{file=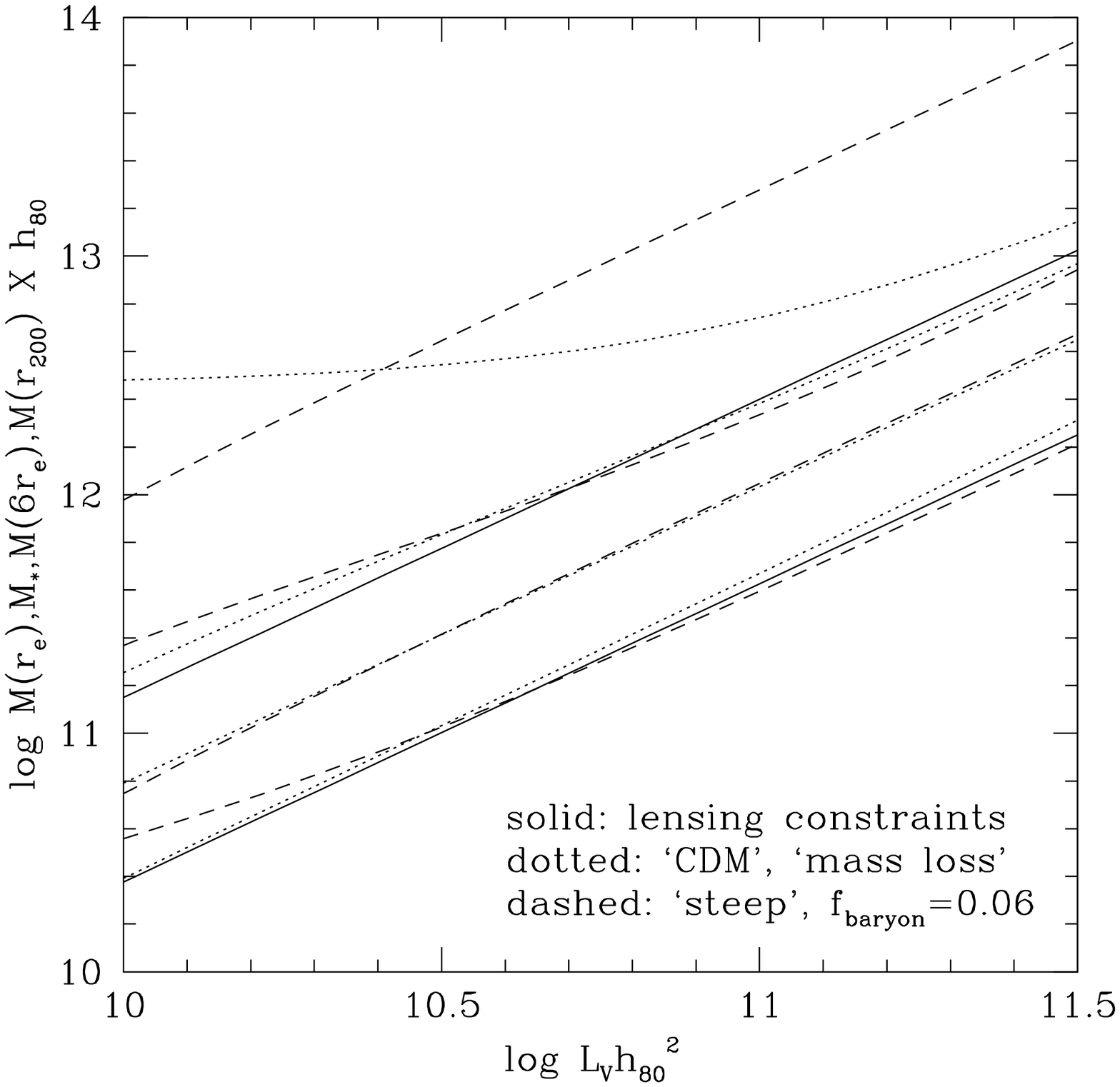,width=2.5in,height=2.5in,clip=}}
\caption{$M$ vs $L_V$ within
(bottom to top) 
$R_e$, $R_{\rm opt}$ (stars), $R_{\rm opt}$ (total), and $R_{\rm virial}$.
Solid curves show 
$M(R_e)$ and $M(R_{\rm opt})$ inferred from statistical weak lensing.}
\end{minipage}
\hspace{\fill}
\begin{minipage}[t]{65mm}
\centerline{
\psfig{file=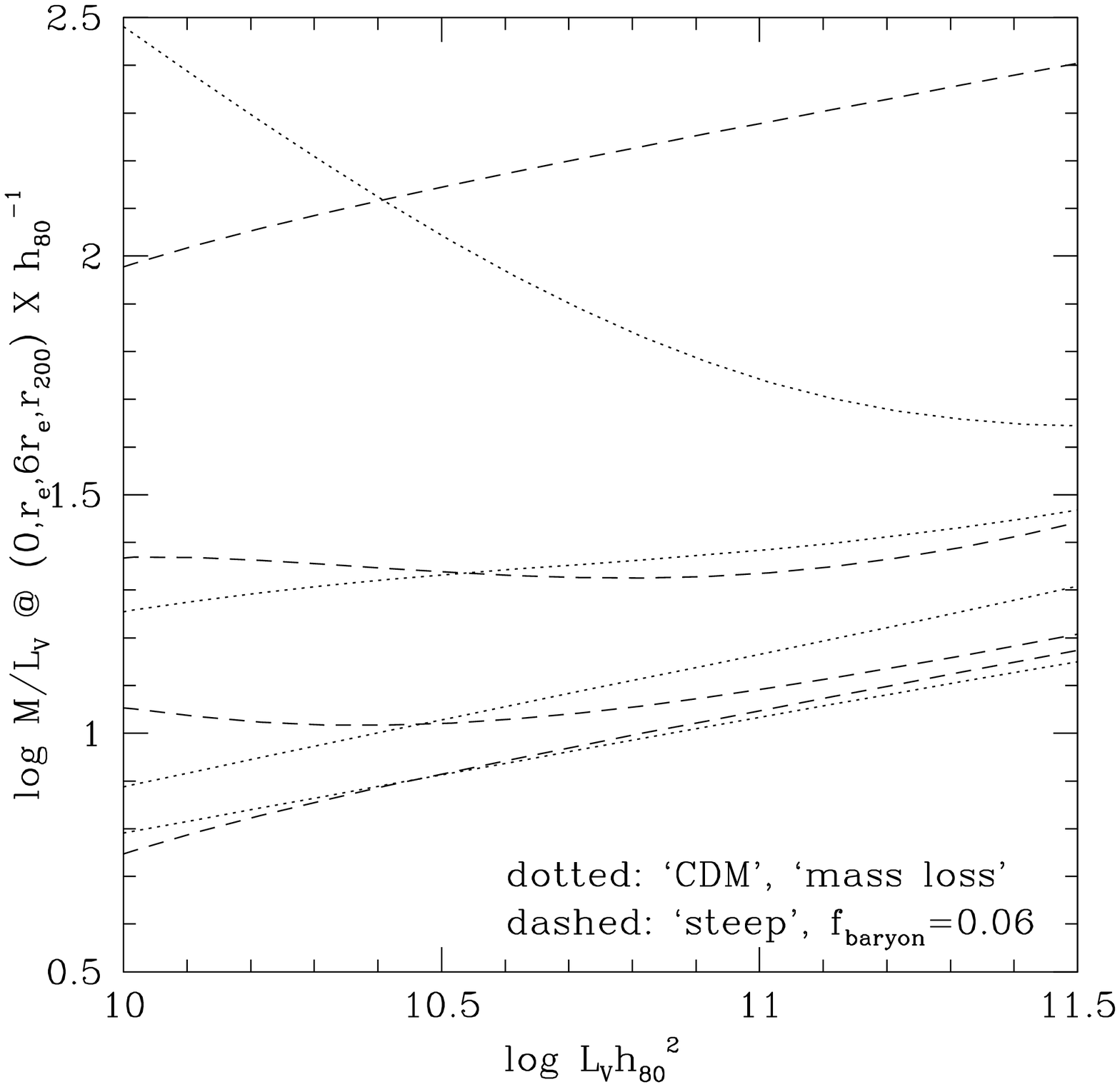,width=2.5in,height=2.5in,clip=}}
\caption{Same as Figure 9 for $M/L_V$ at
(bottom to top): $r=0, R_e, R_{\rm opt}, R_{\rm virial}$.}
\end{minipage}
\end{figure}

We have calculated dark matter and
stellar
velocity dispersion distributions, assuming isotropic orbits.
These are compared in Figure 11 for an 
$L_V=5.2\times 10^{10}{h_{80}}^{-2}$L$_{V_\odot}$ galaxy for both of
the successful scenarios described above and in Figures 9 and 10. 
Both distributions have maxima since the total gravitational
potential is not isothermal. The ratio 
(dark-matter-to-stars) of the squares of these maxima
is greater than 1.4
over the luminosity range in Figures 9 and 10,
and is $\approx 2$ over the range
$L_*<L_V<5L_*$.
In fact, the minimum value of this ratio for any model 
that produces $\beta_{\rm spec}<0.7$ is greater than one. In this sense
the dark matter is hotter than the stars, as simply reflected by the
observation that the gas temperature exceeds that of the stars.

\begin{figure}[htb]
\centerline{
\psfig{file=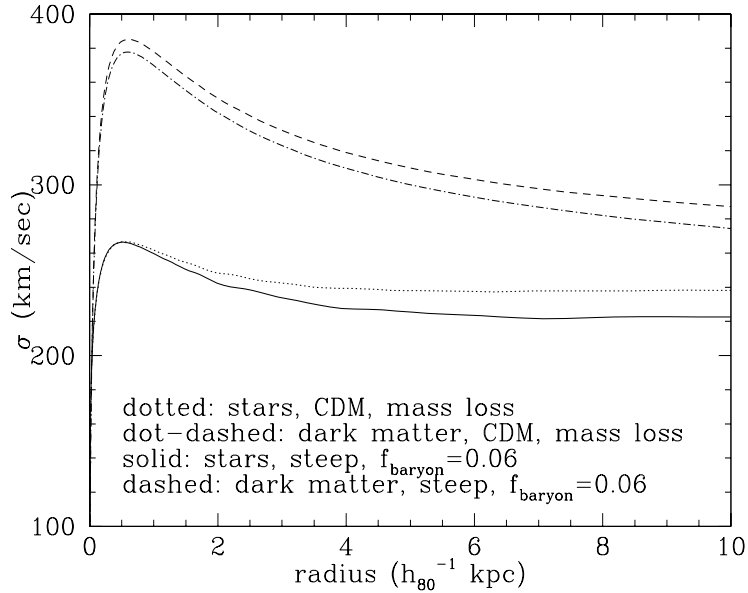,width=4.0in,height=3.2in,angle=-90,clip=}}
\caption{1-d velocity dispersion distributions, assuming isotropic orbits,
for an $L_V=L_0=5.2\times 10^{10}{h_{80}}^{-2}$L$_{V_\odot}$ galaxy.}
\end{figure}

\section{Summary}

This review has focussed on two major investigations of the
hot, X-ray emitting ISM in elliptical galaxies of relevance to the
issue of star formation in elliptical galaxies --  the nature and
metallicity of the hot gas, and the properties of the dark matter
halos confining the hot gas.

\subsection{Abundances in the Hot ISM: Concluding Remarks}

X-ray spectra of elliptical galaxies are adequately fit by models
consisting of hot gas with subsolar Fe abundance and
roughly solar Si-to-Fe ratio, plus
a hard component from an ensemble of X-ray binaries.
The consistency of the magnitude and spectrum
of the hard component with that expected from X-ray binaries and
its compact morphology, support this model over
ones where the hard component is primarily due to a hotter
gas phase. Complications in the form of an extra soft continuum
or multiple phases can be considered, but the consistency of the Si line
diagnostic and continuum temperatures demonstrates that 
the data -- at the present level of sensitivity and spectral resolution --
do not require these.
Optical and X-ray   
Fe abundance estimates are converging, although there are some
cases with anomalously low X-ray values.

Occam's razor would seem to demand that we provisionally
accept the reality of
low abundances in elliptical galaxies. As a result, we need to
seriously
reevaluate our notions of elliptical galaxy chemical
evolution, intracluster enrichment, and Type Ia supernova rates.

\subsection{Dark Matter in Elliptical Galaxies: Main Conclusions}

We (Loewenstein \& White 1998)
have constructed mass models of elliptical galaxies 
consistent with the fundamental plane 
scaling relations and {\it HST} 
results on the structure of the centers of elliptical galaxies, with
dark halos as predicted by
large scale structure
formation simulations. These models allow us to 
calculate the
diagnostic parameter
$\beta_{\rm spec}$ as a function of the relative (to luminous)
dark matter mass and scale length. Comparison with the
observed mean $T$--$\sigma$ relation -- the main features of which are that
the X-ray emitting gas is always hotter than the stars, and by an amount
that increases for
galaxies of lower velocity dispersion/optical luminosity --
provides constraints on the
properties of dark halos around elliptical galaxies. Our main results are as
follows.

(1) In the absence of dark matter, $\beta_{\rm spec}$ generally
exceeds 1.2, with an absolute lower limit of 0.75. Since
galaxies are observed to have $\beta_{\rm spec}=0.3$--0.8, we conclude
that dark halos are generic to $L>L_*$
elliptical galaxies.

(2) The most natural explanation of the 
observed correlation of $\beta_{\rm spec}$
with luminosity is that 
less luminous galaxies are more dark-matter dominated inside $R_{\rm opt}$
in such a way that the total mass-to-light ratio 
is nearly constant. This ratio, 
$\approx 25h_{80}$M$_{\odot}$/L$_{V_\odot}$, is exactly what
is predicted for mass models of elliptical galaxies designed to
explain the gravitational shear of background field galaxies
measured for a disjoint sample of elliptical galaxies.

(3) Our models can be embedded within theories of large scale structure
by specifying how the  dark
matter concentration
scales with virial mass, and linking the virial mass
to the observed luminosity by specifying a global baryon fraction.
The standard CDM scaling with constant baryon fraction badly
fails to reproduce the observed $T$--$\sigma$ relation, since it
predicts an increase in dark-to-luminous ratio (inside $R_{\rm opt}$)
with luminosity.
The following two successful variations are obtained by relaxing one 
of the two assumptions
of the constant baryon fraction CDM scenario:
(a) standard CDM scaling for the dark halos, but with smaller 
galaxies losing an increasingly large fraction of their initial
baryonic content; or,
(b) a constant baryon fraction, but with the
dark-matter concentration 
varying much more strongly with virial mass 
than CDM models predict so that 
more luminous galaxies are less dark-matter dominated due to a
relatively diffuse (rather than less massive) dark halo.
 
\acknowledgments
I am grateful to Scott Trager and Kyoko Matsushita
for providing unpublished results, and to
Richard Mushotzky and Ray White for their collaboration
on this work.
I would also like to thank Patricia Carral and the organizers for
a meeting that was outstanding in every way, and to
Omar Lopez-Cruz for his guidance.

\begin{question}{Vladimir Avila Reese}
Have you included the gravitational pull of the collapsing baryonic matter
on the dark matter halo?
\end{question}
\begin{answer}{Loewenstein}
I've calculated such distortions using the
adiabatic approach of Blumenthal et al., but have not
incorporated the altered halos into our models -- primarily because such an
orderly collapse now seems to be a poor approximation to the actual
formation of ellipticals. Clearly the effects of baryon evolution
on dark halo structure are important and require further study; however,
much depends on the relative timescales of merging, dissipation, 
and star formation.
\end{answer}
\begin{question}{Richard Bower}
Why do you need to assume a model for the dark matter profile? Why not infer
this directly (and uniquely) from the temperature and density profiles?
What uncertainty does this introduce?
\end{question}
\begin{answer}{Loewenstein}
Because our goal in this project was to examine the dark matter properties in
a statistically meaningful sample, we necessarily include galaxies with
only moderately good X-ray data, {\it i.e} where only
a single integrated temperature is derivable and the dark matter profile
{\it cannot} be uniquely determined. For the sake of uniformity
we consider only an average temperature for each galaxy in
the sample, where the
average is taken over an identical {\it metric} radius of $6R_e$.
The total amount of dark matter
within this radius is very well determined, but its detailed distribution is 
not; the NFW function is chosen as a matter of convenience.
We hope to follow this general study up
with a closer look at individual cases where
moderate constraints may be placed on the 
form of the dark matter distribution, although the crude spatial 
resolution of X-ray
temperature profiles imposes severe limitations.
\end{answer}
\begin{question}{Paul Eskridge}
How much is the $T$--$\sigma$ relation of Davis and White effected
by the hard (non-gaseous) X-ray emission from the relatively faint,
low-$L_x$ galaxies?
\end{question}
\begin{answer}{Loewenstein}
Although Davis and White do not include the hard component in their
spectral fits to {\it ROSAT} data, the temperatures they derive
are in superb agreement with {\it ASCA} spectral analysis that
{\it does} include the hard component. I also believe that
the trend is primarily driven by gas-rich galaxies that have the smallest
temperature uncertainties by virtue of their higher luminosities.
\end{answer}
\begin{question}{Paul Goudfrooij}
I think that a significant number of ``normal'' ellipticals exhibiting
X-ray emission from hot gas are dominant members of galaxy groups
(small groups of the Huchra/Geller type), so that their low values of
$\beta_{\rm spec}$ may be partly due to the fact that they reside in the
center of the group, as well as their own, potential. That is, the
hot gas temperature should be compared to the equivalent of the ``combined''
velocity dispersion of the galaxy plus that of the group in which it resides.
Could you comment on this?
\end{question}
\begin{answer}{Loewenstein}
The relevant velocity dispersion for our study
that aims to constrain the dark halo relative to 
optical galaxy properties 
within the (optically) luminous part of the galaxy
is the {\it central} velocity dispersion, since it
is one of the fundamental plane parameters and sets 
the stellar mass scale. The group velocity dispersion becomes of interest if
compared  with 
with the outer temperatures of the very extended X-ray
halos.
\end{answer}
\begin{question}{Michael Pahre}
There is recent evidence that the scaling of velocity dispersion
from central to large radial values may be varying systematically as a 
function of galaxy luminosity (e.g., Busarello et al. 1997 on mergers
of dissipationless systems). How might your results on the
luminosity-dependence of dark matter be effected by this property?
\end{question}
\begin{answer}{Loewenstein}
The relative unavailability of velocity dispersion profiles, as well as
the sort of complications you raise, motivated our exclusive
consideration of
central velocity dispersions in this study. For the more detailed study 
we have planned, whatever dark halo structure we consider must confront
the observations you describe.
\end{answer}
\begin{question}{Daniel Thomas}
Bright ellipticals host $\alpha$-enhanced stellar populations. If it is
mainly these galaxies that enrich the ICM, would you expect a
galaxy/ICM asymmetry (Renzini et al. 1993) in the sense that
Mg/Fe is underabundant in the ICM? {\it ASCA} data point towards ratios
that are at least not subsolar. What do you think is the best way out of this
dilemma?
\end{question}
\begin{answer}{Loewenstein}
Although we have little information on Mg/Fe in the ICM,
ratios of other $\alpha$-elements relative to Fe are supersolar.
The {\it observed} lack of a
galaxy/ICM asymmetry implies that one of the assumptions of
Renzini et al. (1993) -- that star formation in proto-ellipticals has the
same mix of SNIa and SNII as our own Galaxy and that the enrichment process in
the ICM is prolonged compared to that of the stars -- must be
abandoned. Also, I believe Renzini et al. probably
overestimated the total amount of Fe locked up in stars by a factor of
2--3. More puzzling to me is why the Si/Fe ratio in the hot ISM is
$\sim$solar, in apparent conflict with the stellar Mg/Fe ratio.
\end{answer}

\end{document}